\documentclass[runningheads,citeauthoryear]{apinv}
\usepackage{cite,graphics}
\usepackage[utf8]{inputenc}
\usepackage{graphicx}
\usepackage{marvosym}
\usepackage{url}
\usepackage{txfonts}
\usepackage{multirow}
\usepackage{longtable}
\usepackage{adjustbox}
\usepackage{tabularx}
\usepackage{caption}

\usepackage[pagewise]{lineno}

\usepackage{xcolor}

\setkeys{Gin}{draft=False} 

\begin{document}

\title{Modeling Metric Gyrosynchrotron Radio Emission From the Quiet Solar Corona}
\titlerunning{Quiet Coronal Metric Gyrosynchrotron Emission}

\author{\textsuperscript{1}Kamen Kozarev\thanks{Corresponding author} \and \textsuperscript{2}Mohamed Nedal}
\authorrunning{K. Kozarev et al.}

\institute{\textsuperscript{1}Institute of Astronomy and National Astronomical Observatory - Bulgarian Academy of Sciences, 1784 Sofia, Bulgaria
\newline
\textsuperscript{2}Astronomy \& Astrophysics Section, School of Cosmic Physics, Dublin Institute for Advanced Studies, DIAS Dunsink Observatory, Dublin D15 XR2R, Ireland
\newline
\email{kkozarev@astro.bas.bg} 
}
\authorrunning{K. Kozarev et al.}

\maketitle
\begin{abstract}
The radio emission of the quiet Sun in the metric and decametric bands has not been well studied historically due to limitations of existing instruments. It is nominally dominated by thermal brehmsstrahlung of the solar corona, but may also include significant gyrosynchrotron emission, usually assumed to be weak under quiet conditions. In this work, we investigate the expected gyrosynchrotron contribution to solar radio emission in the lowest radio frequencies observable by ground instruments, for different regions of the low and middle corona. We approximate the coronal conditions by a synoptic magnetohydrodynamic (MHD) model. The thermal emission is estimated from a forward model based on the simulated corona. We calculate the expected gyrosynchrotron emission with the Fast Gyrosynchrotron Codes framework by \cite{Fleishman:2010}. The model emissions of different coronal regions are compared with quiet-time observations between 20-90 MHz by the LOw Frequency ARray (LOFAR;\cite{vanHaarlem:2013}) radio telescope. The contribution of gyrosynchrotron radiation to low frequency solar radio emission may shed light on effects such as the hitherto unexplained brightness variation observed in decametric coronal hole emission, and help constrain measurements of the coronal magnetic fields. It can also improve our understanding of electron populations in the middle corona and their relation to the formation of the solar wind.
\end{abstract}
\keywords{Solar Physics, Radio Emission, Quiet-time corona}

\section{Introduction}
\label{introduction}
The solar atmosphere has been well studied in various parts of the electromagnetic spectrum. In particular, with the launch of modern, space-based dedicated solar observatories from the 90s on, there has been an explosion of new imaging observations, mostly in EUV, X-ray, and visible light. These data have allowed researchers to improve our understanding of the processes governing the heating and dynamics of the solar corona. The low solar atmosphere has also been studied with high-frequency radio observations using dish telescopes generating raster images. However, the coronal radio emissions at metric-decimetric frequencies have been mostly explored in connection to solar bursts with solar radiospectrometers. These allow to estimate parameters of energy release in flares and coronal mass ejections (CME), but not to pinpoint the sources of the emission. Thus, the quiet solar corona has not been studied well with low-frequency radio observations. This has recently changed because of the advent of modern, distributed, interferometric arrays with many simple elements and large collecting area, such as the Low Frequency Array (\cite{vanHaarlem:2013}; LOFAR) and the Murchison Widefield Array (\cite{Tingay:2013}; MWA). These new radio telescopes have allowed to create detailed solar images at multiple frequencies between 20 and 300 MHz.

Low-frequency solar imaging is a relatively underexplored domain, yet it holds significant potential for advancing our understanding of the solar corona's structure and activity. Observations in the metric-decimetric range, particularly of gyrosynchrotron (GS) emission, can provide valuable insights into electron populations and coronal plasma conditions. However, GS observations remain rare and are typically associated with eruptive events such as solar flares and coronal mass ejections (CMEs). In particular, studies by \cite{Maia:2000}, \cite{Bastian:2001}, \cite{Pick:2006}, among others, have shed light on the possibility of CME-related GS emission.

Recent work by \cite{Mondal:2020} using the MWA in the 80–300 MHz frequency range highlighted the importance of GS emission during CMEs. Their study revealed multiple regions of GS emission and demonstrated the effectiveness of multi-frequency GS modeling. Despite these advances, low-frequency GS emissions during quiet solar conditions remain poorly studied, leaving a gap in our understanding of non-eruptive coronal processes. Observations with the LOFAR telescope, capable of extending imaging to $~$20 MHz, present an opportunity to explore this under-examined regime and validate existing theoretical models.

Additionally, recent studies (\cite{McCauley:2019}; \cite{Rahman:2019}) have shown frequency-dependent variations in low-frequency imaging. One particularly striking feature is the gradual transition in brightness of coronal holes from dark relative to the rest of the corona at frequencies above about 100 MHz to relatively bright at frequencies below this frequency. These variations have been qualitatively attributed to scattering effects in the corona as well as optical depth effects due to the different origin of high-frequency and low-frequency emission. However, these conclusions have been qualitative only. Interestingly, such effects are absent in synthetic images created using the FORWARD modeling tool \cite{Gibson:2016} that account for thermal and gyroresonance emission. An alternative explanation of the coronal hole brightening effect is that GS emission might play a role in augmenting the overall brightness at frequencies below 100 MHz. Investigating these phenomena using advanced imaging techniques and observational capabilities is critical for a deeper understanding of solar coronal dynamics and electron acceleration processes.

The goal of this work is to introduce our initial approach to modeling the coronal emission and to estimating the effect of GS emission on metric radio interferometric observations of the solar corona during quiet times. We introduce our methods in Section \ref{methods}, and present the results in Section \ref{results}. We summarize our findings in Section \ref{summary}.

\section{Methods}
\label{methods}

The quiet-time solar emission at metric-decimetric wavelengths results from a combination of three main emission mechanisms \cite{Nindos:2020}. The dominant emission mechanism is still the thermal bremsstrahlung from the thermal distribution of coronal electrons, and is ubiquitous throughout the solar atmosphere. Close to and above Active Regions (AR) with strong magnetic fields, gyroresonance emission is produced by Lorentz force-accelerated electrons of suprathermal energies ($10^5$-$10^6$~K). Finally, at mildly relativistic electron energies (few 100s of keV), gyrosynchrotron emission may occur, producing a smooth spectral shape. This mechanism is the dominant source of high-frequency ($>$2~GHz) emission observed in solar flares, but its potential contribution has not been well studied. The recent low-frequency observations of CME-related gyrosynchrotron emission beg the question of whether such emission can be relevant for the quiet-time solar corona. We outline below our initial approach to testing this hypothesis by combining several models of coronal plasma and its radio emission, and comparing them with low-frequency imaging observations from LOFAR.

\subsection{Coronal Plasma Model}
To simulate coronal magnetic and plasma conditions as the basis of our work, we utilized steady-state results from the Magnetohydrodynamic Algorithm outside a Sphere (MAS) coronal model (\cite{Linker:2011}). The MAS model takes as input observations of the radial component of the photospheric magnetic field. It provides a resistive thermodynamic MHD description of the coronal magnetic field and plasma properties between the photosphere and 21 solar radii, under steady-state solar conditions. In our analysis, we use data cubes with model results for individual Carrington rotation (CR) periods, freely available online\footnote{https://predsci.com/mhdweb}. The coronal plasma parameters are used as inputs for our analysis of radio emission.

\subsection{Thermal and Gyroresonance Radio Emission Model}
For generating thermal and gyroresonance radio emission synthetic observations, we have used the FORWARD suite of models \cite{Gibson:2016}. This is an IDL SolarSoft package, which has been developed and improved by the solar community with the goal to perform coronal magnetometry, and in essence allow their user to produce a broad range of observable quantities from a given distribution of coronal magnetic fields and associated plasma. It is very versatile, and may be used to access and compare synthetic observables to real data. FORWARD has been extensively validated and used in prior studies to compare modeled emissions with observations from the Murchison Widefield Array (MWA) (\cite{McCauley:2019}; \cite{Rahman:2019}; \cite{Sharma:2020}). FORWARD includes functionality, which allows to produce radio frequency synthetic images using the MAS MHD steady state description of the coronal plasma. The images can be produced for a particular date by rotating the MAS datacubes according to the Earth's position and integrating plasma quantities along the line of sight. The FORWARD model also generates maps of line-of-sight (LOS) integrated plasma quantities, providing a comprehensive view of the modeled coronal conditions.

\begin{figure}[!htp] 
  \centerline{\includegraphics[width=0.95\columnwidth]{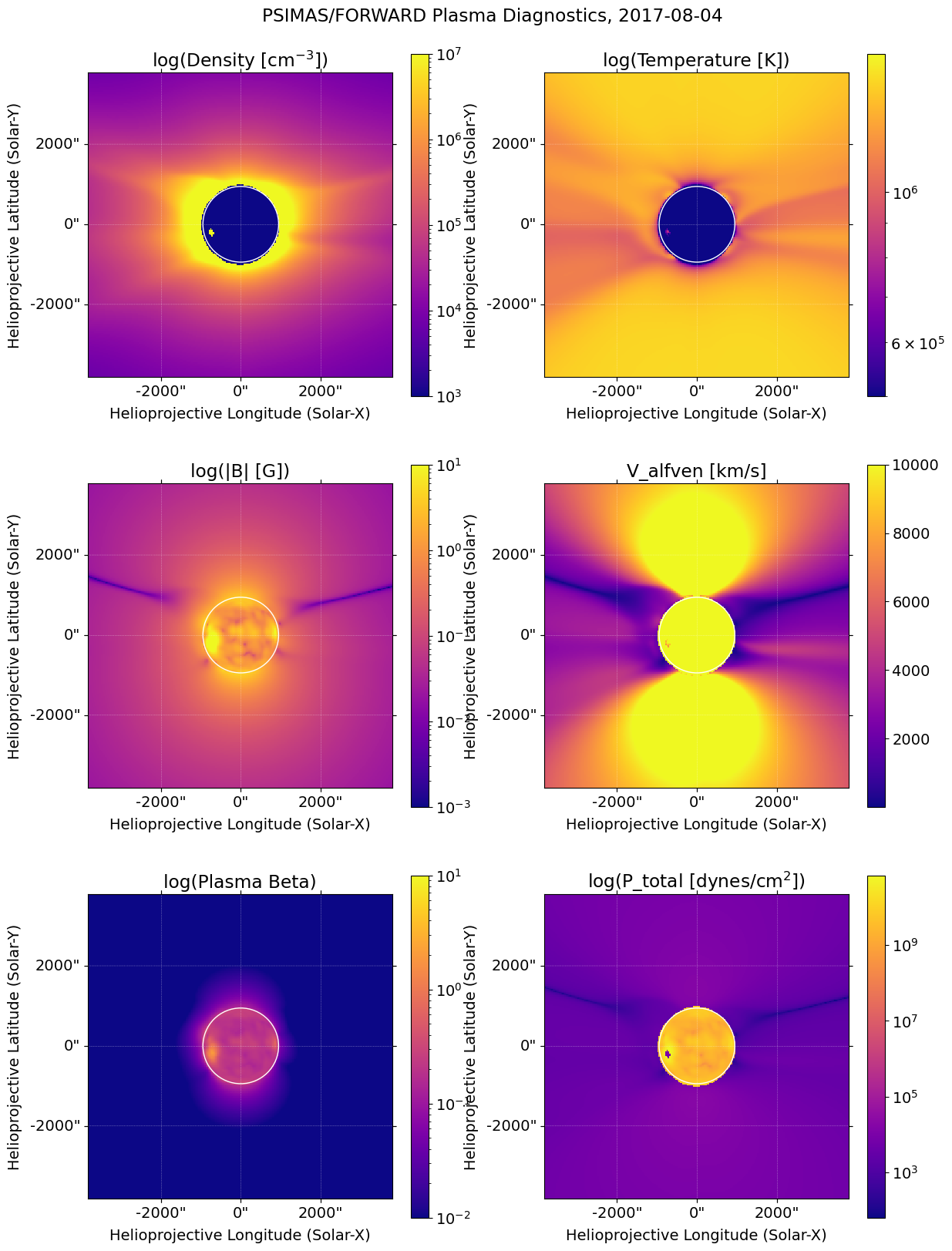}}
  \caption{FORWARD+MAS plane-of-sky maps of coronal plasma quantities for August 04, 2017 (minimum of solar cycle). The white circles denote the extent of the optical solar disk.}
  \label{f01}
\end{figure}

\begin{figure}[!htp] 
  \centerline{\includegraphics[width=0.95\columnwidth]{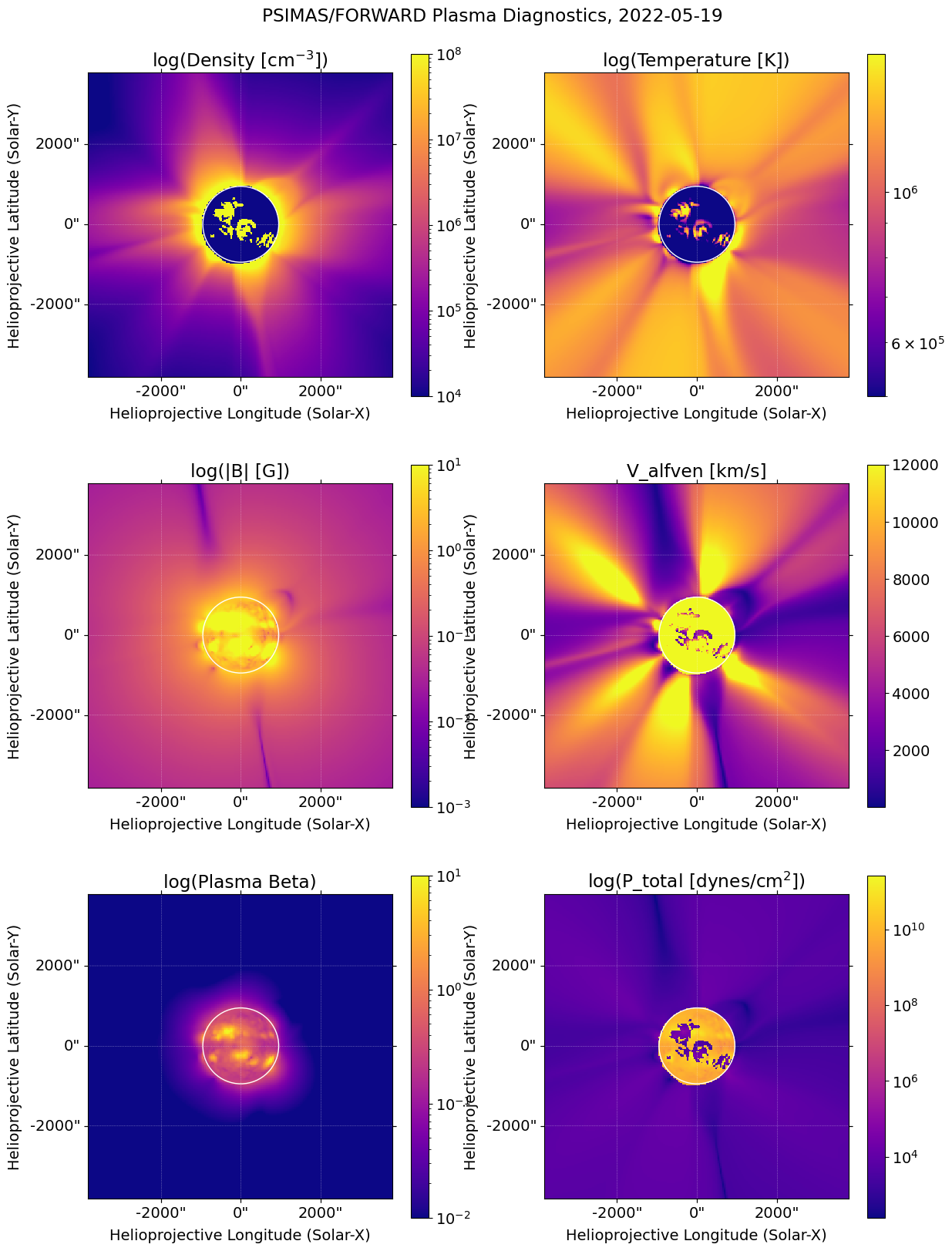}}
  \caption{FORWARD+MAS plane-of-sky maps of coronal plasma quantities for May 19, 2022 (maximum of solar cycle). The white circles denote the extent of the optical solar disk.}
  \label{f02}
\end{figure}

In this study, we used FORWARD to generate thermal bremsstrahlung and gyroresonance emission maps and compared them with LOFAR quiet-time radio observations, for the same fields of view and LOFAR frequency subbands. These comparisons were done for both solar activity minimum (04 August, 2017) and solar maximum (19 May, 2022) coronal conditions. This allowed us to investigate the relationship between modeled thermal emissions and observed low-frequency radio brightness. Figures \ref{f01} and \ref{f02} show the plane-of-sky maps of coronal plasma quantities (number density, temperature, total magnetic field, Alfv\'en speed, plasma beta, and total pressure), generated by FORWARD using the MAS results for the two dates mentioned above. The maps show the clear difference in the structure of the solar corona during the solar minimum (Fig. \ref{f01}) and maximum (Fig. \ref{f02}). Additionally, they shows the spatial resolution of the MHD model, which is more than sufficient for producing detailed synthetic radio images. The thermal images of coronal brightness temperature in Kelvin, produced by FORWARD, are shown in the middle-right panels of Figures \ref{f03} and \ref{f04}. These were then convolved with a synthetic point-spread function (PSF), determined from the LOFAR imaging routines, and shown on the bottom right panels of Figs. \ref{f03} and \ref{f04}, in order to make direct comparison with the actual observations. The resulting final images are shown on the top right panels.

\subsection{LOFAR radio observations}
The interferometric imaging of the quiet corona for this work was carried out using low-band antenna (LBA) data from LOFAR’s Core and Remote Stations, enabling less than arcminute-scale spatial resolution. The interferometric data processing follows a series of steps. Initially, gain calibration is performed using the DP3 (Default Pre-Processing Pipeline)\cite{vanDiepen:2018} code on Cas-A and Sgr-A calibrator observations, which generates phase and amplitude corrections based on a reference flux density model. These gain solutions are computed for each baseline and subband. The derived calibration solutions are then applied to the target observations. Imaging is done by converting visibility data from wave vector space (u, v) to image space (x, y) using a 2D Fourier Transform with the w-stacking CLEAN algorithm implemented in wsclean code \cite{Offringa:2014}. In the final post-processing step, images are converted to helioprojective coordinates and units are translated into brightness temperature for scientific analysis using the lofarSun \cite{Zhang:2022} Python toolset. 

\begin{figure}[!htp] 
  \centerline{\includegraphics[width=0.95\columnwidth]{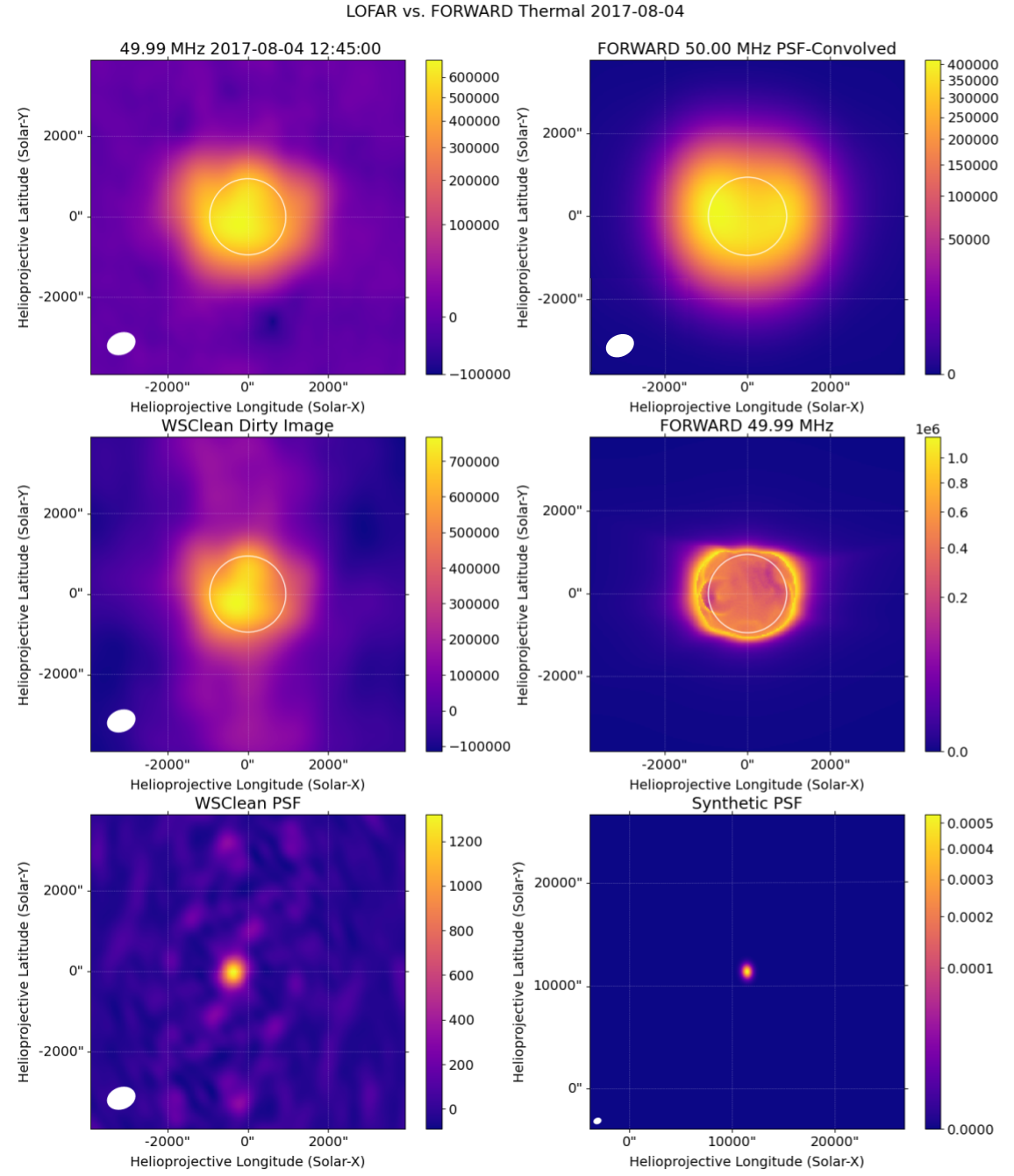}}
  \caption{LOFAR interferometric image (top, left) at 50 MHz and synthetic FORWARD emission for August 04, 2017 (top, right). Middle rows show the dirty LOFAR image (left) and non-PSF-convolved FORWARD image (right), respectively. Bottom rows show the shapes of the beams/PSFs used for the LOFAR image generation (left) and the convolution of the FORWARD image (right). The white circles denote the extent of the optical solar disk.}
  \label{f03}
\end{figure}

\begin{figure}[!htp] 
  \centerline{\includegraphics[width=0.95\columnwidth]{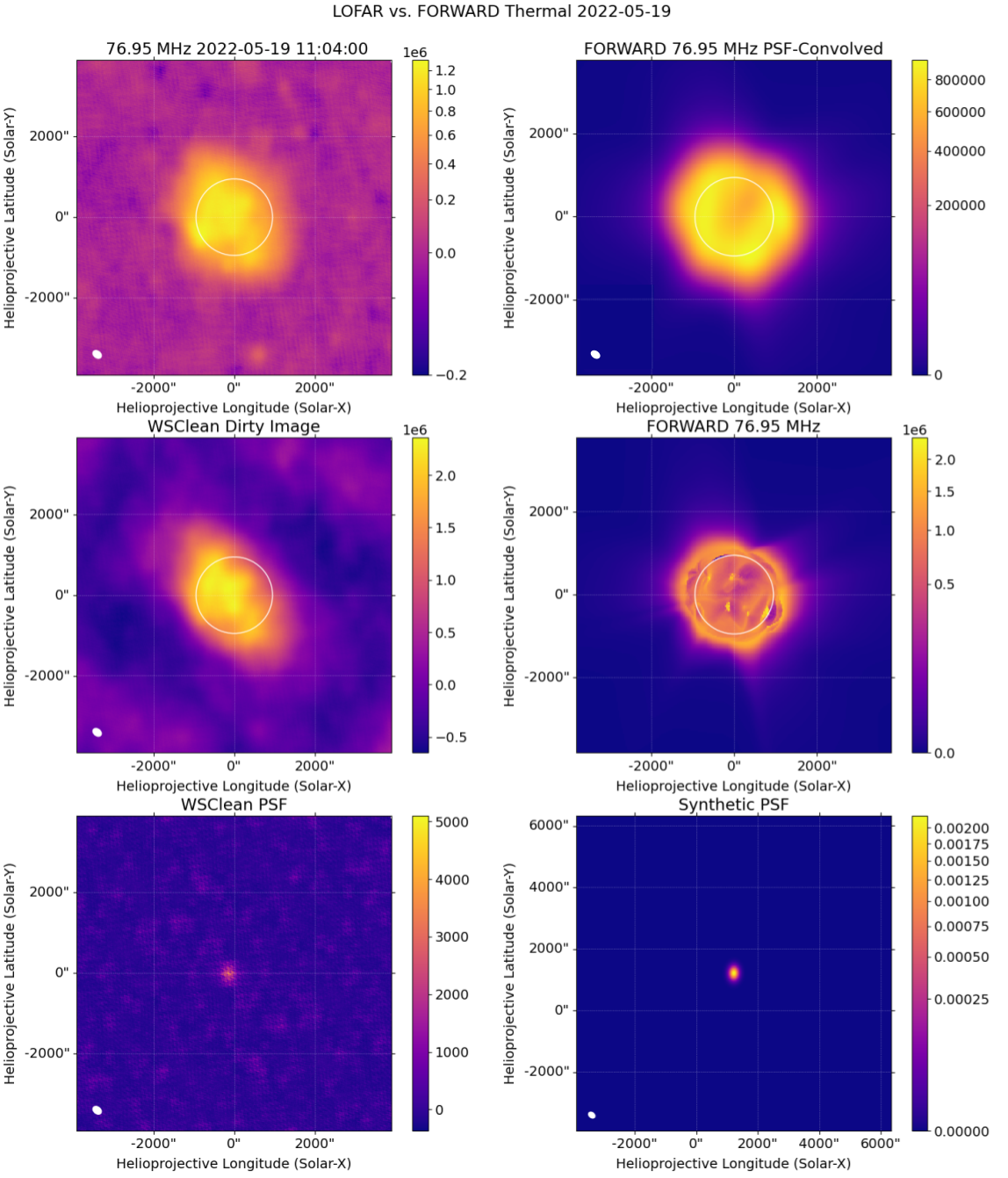}}
  \caption{LOFAR interferometric image (top, left) at 77 MHz and synthetic FORWARD emission for May 19, 2022 (top, right). Middle rows show the dirty LOFAR image (left) and non-PSF-convolved FORWARD image (right), respectively. Bottom rows show the shapes of the beams/PSFs used for the LOFAR image generation (left) and the convolution of the FORWARD image (right). The white circles denote the extent of the optical solar disk.}
  \label{f04}
\end{figure}

\begin{figure}[!htp] 
  \centerline{\includegraphics[width=0.99\columnwidth]{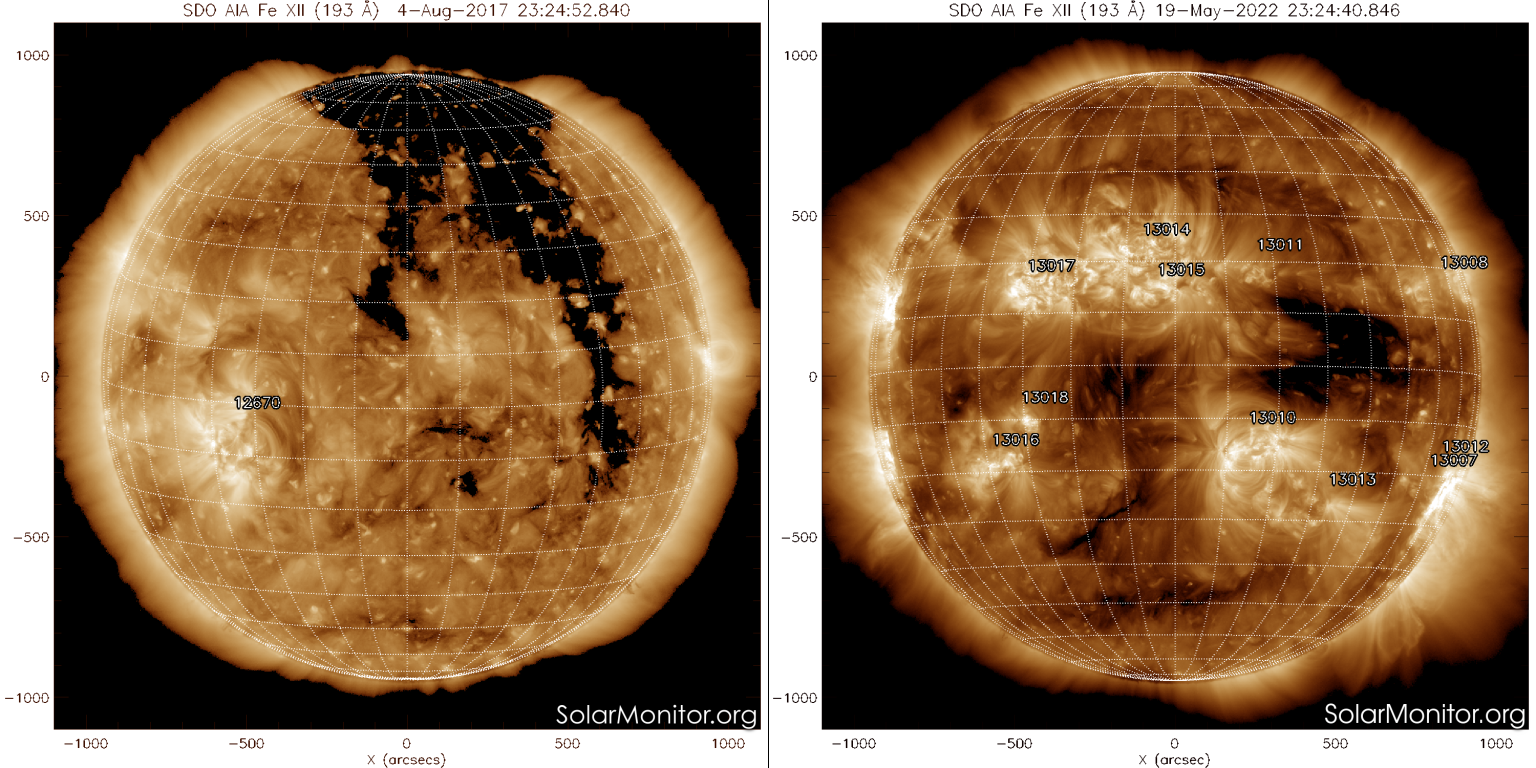}}
  \caption{Extreme UV observations by the SDO/AIA instrument in the 193\AA~channel show the structure of the low solar corona on August 08, 2017 (left) and May 19, 2022 (right).}
  \label{faia}
\end{figure}

The processing steps are shown in the left panels of Figures \ref{f03} and \ref{f04} for August 04, 2017 at 49.99 MHz and May 19, 2022 at 76.95 MHz, respectively. The gridded Fourier transforms of the visibilities (so-called 'dirty images') are shown in the middle left panels. The LOFAR effective 2D PSFs are shown in the bottom left panels, and the final ('cleaned') images are shown on the top left panels. The coronal structure is clearly visible, showing bright emission spots (corresponding to active regions) on the solar disk and large-scale off-limb diffuse emission, corresponding to streamers. Figure \ref{faia} shows for context extreme ultraviolet (EUV) images on the two dates in the 193~\AA channel of the Solar Dynamics Observatory's Atmospheric Imaging Assembly space telescope. Comparing the EUV image pairs, the difference in activity phase is quite obvious, with 11 active regions identified in the 2022 image versus one AR dominating the southeast disk in the 2017 one. The on-disk radio emission in the May 19, 2022 LOFAR image also shows a number of bright regions, but it is difficult to assign them directly to their EUV counterparts. As expected, the large coronal hole visible in the August 04, 2017 EUV image is not seen in the LOFAR image on Fig. \ref{f03}.

\subsection{Modeling of Gyrosynchrotron Emission}
\begin{figure}[!htp] 
  \centerline{\includegraphics[width=\columnwidth]{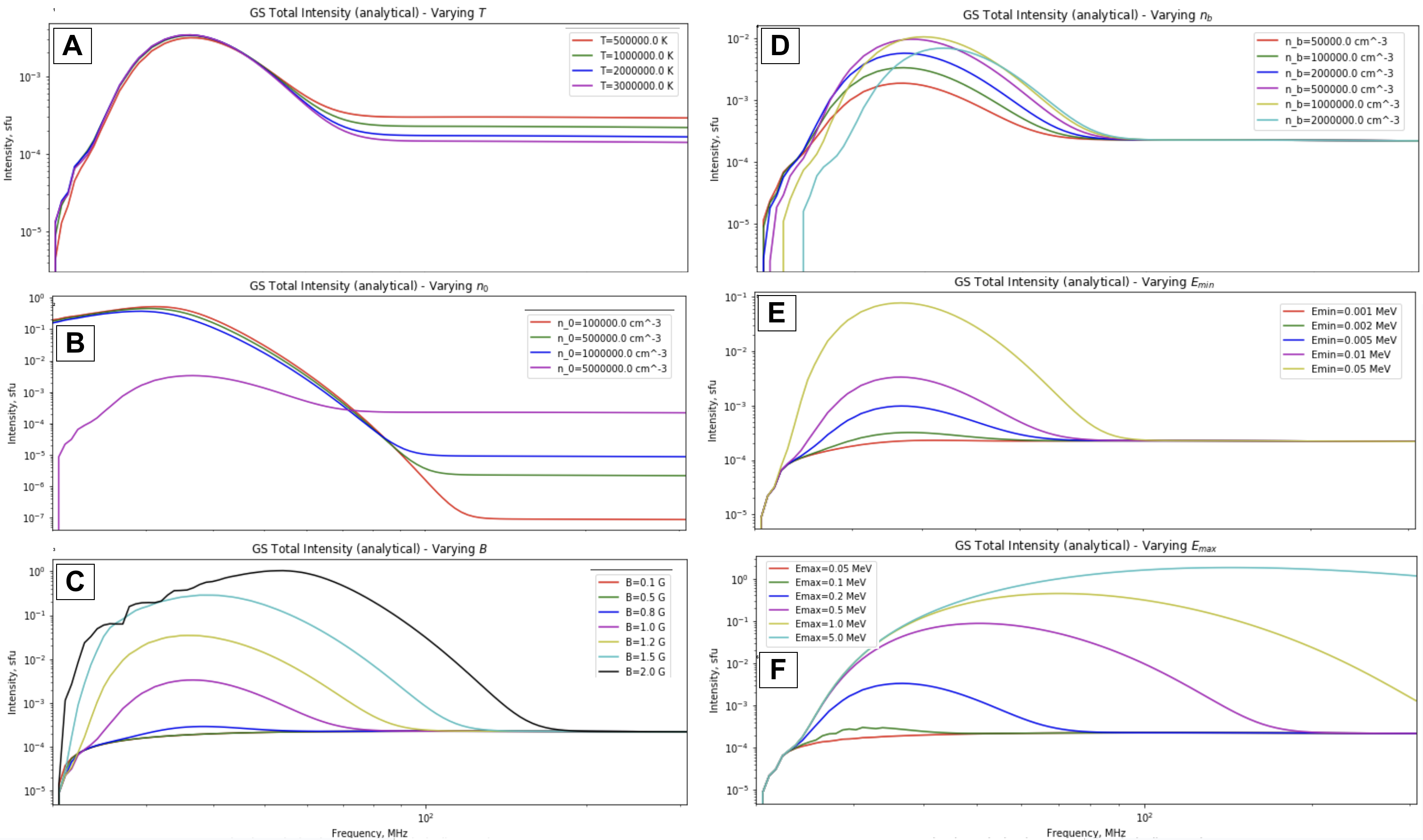}}
  \caption{Testing low-frequency GS spectra between 20-300 MHz, varying the main model parameters according to Table \ref{Table1}.}
  \label{f05}
\end{figure}

For the modeling of gyrosynchrotron emission, we used the Fast GS Codes simulation suite \cite{Fleishman:2010}; \cite{Kuznetsov:2021}. It was initially developed for modeling high-frequency (microwave) solar flare radio emissions. However, it has also been applied to gyrosynchrotron (GS) emission fitting in studies of low-frequency coronal radio emissions, particularly in the Murchison Widefield Array (MWA) range of 80–300 MHz. The main assumption of the model used here is that an electron beam with a power-law distribution is responsible for the emission. The inputs of the model are coronal temperature, thermal plasma density, magnetic field strength, non-thermal electron density, as well as the minimum and maximum energies of the power law electron distribution. In this study, we extended the application of the Fast GS Codes to explore the GS emission domain at the lowest observable frequencies. To achieve this, we designed a grid of parameters representative of the low-to-middle corona during a quiet period (Table \ref{Table1}), aligning with conditions described in previous studies, such as those summarized by \cite{Mondal:2020}. The table shows the various combinations of the input parameters to the FastGS code, with letters A-F corresponding to the resulting model spectra, shown in Figure \ref{f05}. This approach allowed us to investigate 1) whether significant GS emission is possible at these frequencies, 2) what the characteristics of GS emission at these frequencies are, and 3) assess its relevance to observed low-frequency coronal radio emissions.

\begin{center}
\begin{longtable}{ccccccc}
\caption{Baseline parameter values and variation ranges for the GS gyrosynchrotron codes model used in the simulation of slow solar wind from the interfaces between coronal streamers and holes.}\\
Panel	&$T_0$	&$n_0$	&$B$	&$n_b$	&$E_{min}$	&$E_{max}$\\
	&(K)	&($cm^{-3}$)	&(G)	&($cm^{-3}$)	&(MeV)	& (MeV)\\
\hline\endhead
\hline
A & \textbf{[1.e5-1.e7]} & 5.e6 & 1. & 1.e5 & 0.01 & 0.2\\
B & 3.e6 & \textbf{[1.e5-5.e7]} & 1. & 1.e5 & 0.01 & 0.2 \\
C & 3.e6 & 5.e6 & \textbf{[0.1-1.5]} & 1.e5 & 0.01 & 0.2 \\
D & 3.e6 & 5.e6 & 1. & \textbf{[5.e4-2.e6]} & 0.01 & 0.2 \\
E & 3.e6 & 5.e6 & 1. & 1.e5 & \textbf{[0.001-0.05]} & 0.2 \\
F & 3.e6 & 5.e6 & 1. & 1.e5 & 0.01 & \textbf{[0.05-5.0]} \\
\hline
\label{Table1}
\end{longtable}
\end{center}

The results of the GS modeling in Fig. \ref{f05} show that there is indeed significant gyrosynchrotron emission possible at frequencies between 20 and 100 MHz under the realistic coronal conditions in Table \ref{Table1}. Most of the models we ran produced coronal brightness 10$^-3$-1 solar flux units (sfu), except for the lowest magnetic fields and non-thermal energy ranges.These positive results gave us confidence to continue developing a technique for modeling low-frequency GS emission in the corona. To create synthetic GS images, we ran grids of Fast GS codes models with 3000 combinations of [n,T,B] each, between 20-300 MHz with 100 frequency points. We saved results to look-up tables, and created functions for computing linear interpolation between input plasma values. Grid runs were performed for nine separate combinations of non-thermal electron density (n$_b$), minimum ($E_{min}$) and maximum ($E_{max}$) beam energy. In order to construct GS synthetic images, we sampled the [n,T,B] values in the FORWARD plasma quantity maps (shown in Figs. \ref{f01} and \ref{f02}) pixel by pixel. For each pixel, we extracted the spectrum corresponding most closely to the plane-of-sky [n,T,B] values, and recorded the GS intensity for a desired frequency on the GS image array. This approach has two main limitations, which we hope to overcome in future work. First, using a plane-of-sky plasma value equals the assumption that the coronal plasma does not change along the line of sight, which is obviously not correct. Second, the images are only generated for a given combination of [n$_b$, $E_{min}$, $E_{max}$], which does not change for different locations on and off the solar disk. We expect that these parameters should change in different areas of the solar surface and corona, for example active regions versus coronal holes. We will relax this assumption in future work as well.

\section{Results}
\label{results}

\begin{figure}[!htp] 
  \centerline{\includegraphics[width=0.95\columnwidth]{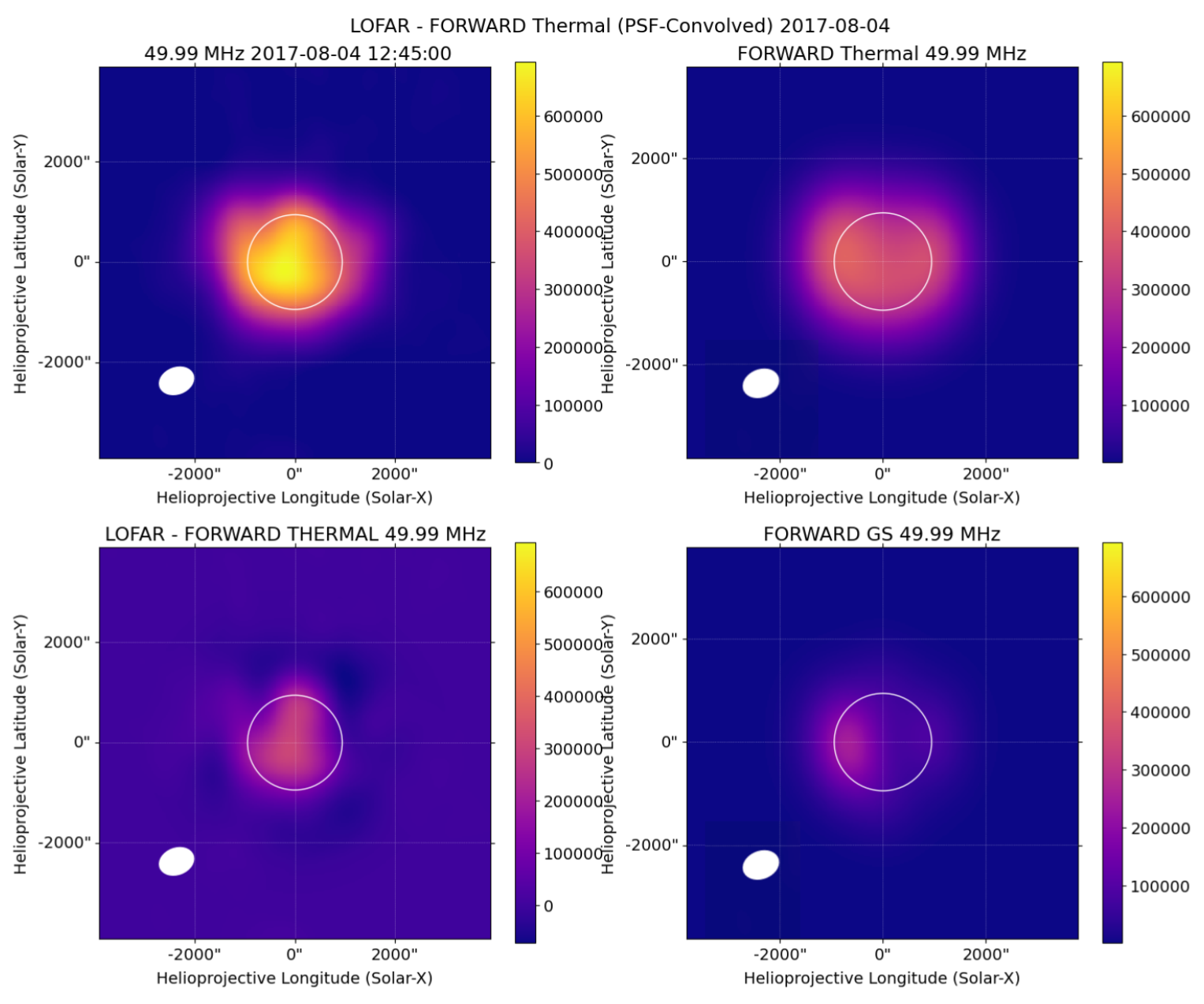}}
  \caption{LOFAR  observations (top left), FORWARD thermal synthetic image (top right), LOFAR - Thermal synthetic (bottom left), and GS (bottom right), for August 4, 2017. The white circles denote the extent of the optical solar disk.}
  \label{f06}
\end{figure}

\begin{figure}[!htp] 
  \centerline{\includegraphics[width=0.95\columnwidth]{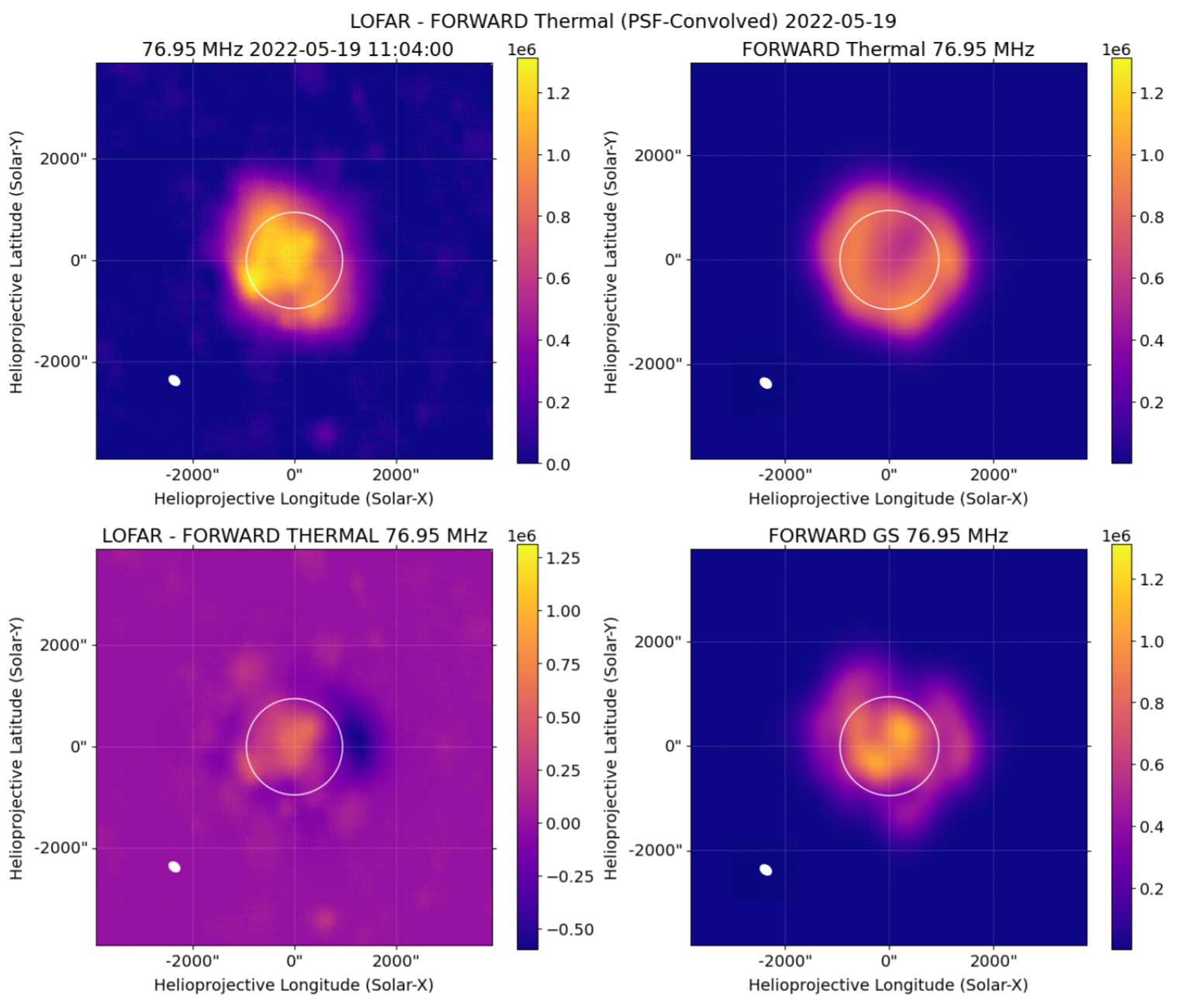}}
  \caption{LOFAR observations (top left), FORWARD thermal synthetic image (top right), LOFAR - Thermal synthetic (bottom left), and GS (bottom right), for May 19, 2022. The white circles denote the extent of the optical solar disk.}
  \label{f07}
\end{figure}

The top panels of Figures \ref{f06} and \ref{f07} provide, for the same radio frequency, a direct comparison between the low-frequency corona observed with LOFAR (left images) and the synthetic coronal images generated with FORWARD under an assumption of thermal and gyroresonance emission only. The two sets of images look different, which can be attributed to several causes. One is the fact that coronal scattering effects have not been taken into account in the FORWARD calculations. This is likely the cause of the well-pronounced limb brightening in the synthetic images, which is not very pronounced in the LOFAR observations. Another possibility is incorrect modelling of the coronal plasma. However, the MAS model produces quite accurate and high-resolution global maps of the corona, as can be seen in Figs. \ref{f01} and \ref{f02}; thus, we do not consider it a major source of discrepancy. Finally, there is the possibility that an additional emission mechanism, namely gyrosynchrotron, may contribute to the overall observed emission. We have set out to test this last possibility, and present our initial results below.

In order to quantify the discrepancy between the observed and synthetic radio corona at low frequencies, we first computed the difference between the LOFAR observations and the FORWARD synthetic images. This was possible since we had taken care to construct the synthetic images with the same field of view and spatial resolution as the LOFAR images. The resulting excess images are shown in the bottom left panels in Figs. \ref{f06} and \ref{f07}. It can be seen that, in both cases, there is still significant emission remaining, mostly concentrated in the central part of the solar disk. We suggest that in both cases a lot of this excess emission comes from active regions, as can be inferred by comparison with the synthetic GS images (bottom right panels of Figs. \ref{f06} and \ref{f07}). 

For August 04, 2017 (Fig. \ref{f06}, bottom right panel), most of the GS emission originates towards the eastern limb of the Sun, which is related to the stronger magnetic fields and the AR there. However, there is a strong central component in the central part of the solar disk in the excess image (Fig. \ref{f06}, bottom left panel), which may correspond to the large coronal hole. There is no corresponding strong emission in the GS image for the input parameters used. This will be investigated further. In the case of May 19, 2022, (Fig. \ref{f07}, bottom right panel), the GS emission from the solar disk is over the active regions. The bright regions at the southeastern and northwestern parts of the solar disk are comparable in size and brightness on both observed and modeled images. The main difference is the off-limb emission, predicted in the synthetic images.

\section{Summary}
\label{summary}

In this work, we present our first approach to investigating the relevance of gyrosynchrotron emission in low frequency radio observations of the solar corona with the LOFAR telescope. We developed a technique for generating synthetic GS images of the solar corona, based on the FastGS code suite and MHD model results. We find that this technique produces meaningful GS images of brightness temperature, comparable to the excess emission in LOFAR images when the thermal synthetic images are subtracted. We find that there is significant excess emission, which is not explained by synthetic thermal and gyroresonance emissions. We propose that gyrosynchrotron emission from active regions in the quiet corona may contribute significantly to explain it. However, the modeling used to generate synthetic GS images includes several simplifying assumptions, which may have a big effect on the results, and which we will strive to relax in further developing the technical routines.

\section*{Acknowledgments}
This work is supported by the Bulgarian National Science Fund, VIHREN program, under contract KP-06-DV-8/18.12.2019 (MOSAIICS project) and by the project "The Origin and Evolution of Solar Energetic Particles”, funded by the European Office of Aerospace Research and Development under award No. FA8655-24-1-7392. The authors acknowledge using data from the Atmospheric Imaging Assembly instrument onboard the Solar Dynamic Observatory, The GS Codes and FORWARD simulation suites, as well as the Sunpy Python library. The authors thank the referee for their valuable feedback.

\bibliographystyle{apalike}
\bibliography{low_frequency_gs_model_BAJ}

\end{document}